\documentclass[oldversion]{aa}  

\usepackage{graphicx}

\newcommand{\CELL}[1]{#1}
\newcommand{\text}[1]{\mbox{#1}}
\newcommand{\RD}[5]{#1 \\}
\newcommand{\RDLAST}[5]{#1 }
\newcommand{\EQN}[5]{\begin{eqnarray} #5 \end{eqnarray}}
\newcommand{\BF}[1]{{\mathbf #1}}
\newcommand{\BFS}[1]{\mbox{\boldmath $#1$}}

\begin{document}
   \title{Radio pulses from cosmic ray air showers}

   \subtitle{Boosted Coulomb and \v{C}erenkov fields}

   \author{N. Meyer-Vernet\inst{1}, 
               A. Lecacheux
\inst{1}, 
\and
D. Ardouin\inst{2}
          }
   \offprints{N. Meyer-Vernet}

   \institute{LESIA, Observatoire de Paris, CNRS, UPMC, Universit\'{e} Paris Diderot; 5 Place Jules Janssen, 92190 Meudon, France\\
              \email{nicole.meyer@obspm.fr}
    \and
SUBATECH, IN2P3/CNRS, Universit\'{e} de Nantes, Ecole des Mines de Nantes, and Observatoire de Paris; 4 rue Alfred Kastler, 44307 Nantes, France
}

   \date{Received September 28, 2007; accepted December 5, 2007.}

 
  \abstract{High-energy cosmic rays passing through the Earth's
atmosphere produce extensive showers whose charges emit radio
frequency pulses.  Despite the low density of the Earth's
atmosphere, this emission should be affected by the air
refractive index because the bulk of the  shower particles move
roughly at the speed of radio waves, so that the retarded 
altitude of emission, the relativistic boost and the emission
pattern are modified. We consider in this paper the contribution
of the   boosted Coulomb and the \v{C}erenkov fields and
calculate analytically the spectrum using a very simplified model
in order to highlight the main properties. We find that typically
the lower half of the  shower charge energy distribution produces
a boosted Coulomb field, of amplitude comparable to the levels
measured and to those calculated previously for synchrotron
emission. Higher energy  particles produce instead a
\v{C}erenkov-like field, whose amplitude may be smaller because
both the negative charge excess and the separation between
charges of opposite signs are small at these energies.}
  

   \maketitle
\section{Introduction}

Nearly a century after the discovery of cosmic rays, their nature
and origin still constitute one of the great problem of
contemporary astrophysics. Most of them are revealed by their
electromagnetic radiation, but those reaching the solar system
play a unique role since they can be observed  nearly directly.
The extensive air showers made of secondary (or higher order)
particles produced by high-energy cosmic rays in the Earth's
atmosphere are currently studied with giant ground-based particle
 detectors, whereas additional information is provided by
observing fluorescence and \v{C}erenkov radiation in the optical
range.

A complementary technique is being developed, based on the
detection of radio emission from the shower charged particles,
making advantage of the techniques of increasing sophistication
developed for radioastronomy. Indeed, it has been known for
several decades that the charges in extensive cosmic ray showers
produce radio frequency pulses, as first suggested by Askar'yan  
(\cite{ask62}) who estimated the \v{C}erenkov emission.
Subsequent studies (see Kahn \& Lerche \cite{kah66})
 have suggested radio emission to be produced instead by
acceleration by the Lorentz force in the Earth's magnetic field
(see the extensive review by Allan  \cite{all71}), and have
stimulated extensive calculations and modelling of the
synchrotron radiation (see Huege \& al. \cite{hue07} and refs.
therein). Up to now, the difficulties in both theory and
observation have precluded a full understanding of the origin of
the radio emission.

Most previous studies of this radio emission have approximated
the air refractive index by unity. However, the median energy of
electrons in the shower is about 30 MeV, corresponding to  
$1-v/c\simeq 1.5\times 10^{-4}$; the median speed is thus greater
than the phase speed $c/n$ of radio waves in air at sea level
(where $n\simeq 1.0003$),  and is roughly equal to the phase
speed at an altitude of half the atmospheric scale height. The
air refractive index may thus  affect significantly the radio
electric field produced by the shower, whatever the emission
mechanism, by changing the retarded altitude, the relativistic
boost and the radiation pattern.

In particular, this should increase the relativistic boost of the
Coulomb field produced by the shower charges, at speeds slightly
smaller than the wave phase speed, whereas faster  charges, which
can catch up with the waves they emit, produce a \v{C}erenkov 
like field. We make below an analytical calculation of these
fields with a highly simplified model of the charge distribution,
in order to highlight the physical processes and evaluate the
contributions to the total electric field. 

Before taking the refractive index into account, let us estimate
the boosted Coulomb field produced in vacuum by a shower
particle. Consider a charge $q$ in uniform relativistic motion.
For an observer at rest, the Coulomb field is radial relative to
the instantaneous present position of the charge, but the
FitzGerald-Lorentz contraction compresses the field lines, so
that in the directions normal to the velocity, the field in
vacuum is greater than the isotropic Coulomb field by the Lorentz
factor $\gamma $. This produces a pulse of electric field of
maximum amplitude $E_{Max}=\gamma q/4\pi \epsilon _0d^2$ when the
charge's path passes at closest distance $d$ to the antenna, and
of temporal width $\tau = d/(\gamma c)$, directed from the
charge's present position to the observation point. The number of
charges in the shower at maximum development is about one per GeV
of primary energy with about  20\% relative excess of electrons
over positrons.  With a primary cosmic ray of  $ 2 \times
10^{17}$eV, i.e. $2\times 10^8$ charges, and a median secondary
energy of 30 MeV ($\gamma =60$), this yields a pulse of maximum
amplitude a few $ 10^{-4}$ V/m   and of duration roughly 10 ns at
100 m perpendicular distance.  The corresponding low-frequency
Fourier transform $ 2 \tau E_{Max} = q/(2 \pi \epsilon _0 dc) $
is a few $\mu $V/m/MHz at frequencies smaller than $1/2  \pi \tau
\simeq 50$ MHz. The exact strength and polarisation are
determined by the magnetic charge separation which produces
oppositely directed changes of position and velocity for charges
of opposite signs. This field should be added to the contribution
of  synchrotron emission.

Since the above amplitude is of the same order of magnitude as
the values observed by different instruments (see Allan
\cite{all71}, Ardouin  \cite{ard06}, Horneffer \cite{hor06}),
this contribution may not be negligible. We therefore consider
this field in more detail, taking the refractive index into
account.

\section{Electric field spectrum of a subluminal charge}

In this Section, we consider a point charge. A charge
distribution and the corresponding coherence effects  will be
considered in Section 4.

\subsection{Retarded potentials}

Consider a point charge $q$ moving in a medium of constant
refractive index $n$ (and relative magnetic permeability $\mu
\simeq 1$), at velocity $\BF{v}$. The field potentials at
space-time point ($\BF{r},t$) are the standard 
Li\'{e}nard-Wiechert potentials with the formal replacements
$c\rightarrow c/n$ and $\epsilon _0\rightarrow \epsilon _0n^2$  
\EQN{7}{1}{}{}{ \RD{ \CELL{\Phi ( \BF{r},t)&= &\frac 1{4\pi
\epsilon _0n^2}\left[ \frac q{\mid 1-n \BFS{\beta}\cdot 
\BF{n}\mid R}\right] _{ret}\label{LW}}}{1}{}{}{} \RDLAST{ \CELL{
\BF{A}( \BF{r},t)  &= & n^2 \BFS{\beta}\Phi (
\BF{r},t)/c\label{LWA}}}{1}{}{}{}}
 where $\BFS{\beta}=\BF{v}/c$,  $\BF{n}$  is the unit vector in
the direction of the point of observation from the moving charge,
$R$ is the distance to the moving charge, and the subscript
``ret'' on the square bracket indicates that the quantities
inside are to be evaluated at the (observer's) retarded time
$t_{ret}=t-nR(t_{ret})/c$ (see for example Feynman  \cite{fey64},
Jackson  \cite{jac99}, Thid\'{e} \cite{thi97}).

Equations  (\ref{LW})-(\ref{LWA}) hold whatever the charge's
motion and the refractive  index, provided $n$ is a constant.
Namely, this formulation neglects both the variation of $n$ with
position and the dispersion, in particular the variation of $n$
with frequency (Ginzburg \cite{gin89}, Clemmow  \& Dougherty
\cite{cle69}). In this paper we will apply these  equations  for
a uniform velocity, so that the electric field, given by 
(\ref{EphiA}), is the acceleration-independent term of the
Li\'{e}nard-Wiechert electric field (see for example Feynman 
\cite{fey64}, Jackson  \cite{jac99}), i.e. mainly a boosted
Coulomb field for a subluminal charge or a \v{C}erenkov-like
field for a supraluminal charge. This field should be added to
the synchrotron field calculated by the current models of shower
emission. In the present Section, we consider the first case:
$\BF{v}$  smaller than the velocity of light in the medium $c/n$,
i.e. $n\beta <1$.

  \begin{figure}
   \centering
 \includegraphics[width=8cm]{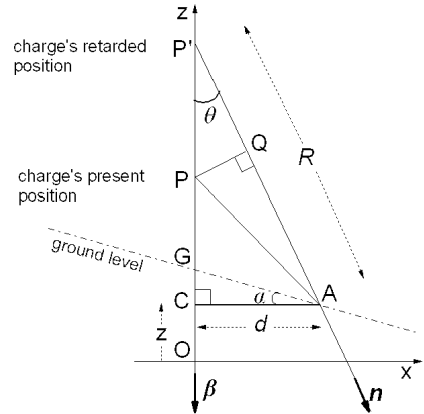}
      \caption{Present (P) and retarded (P') positions of the
moving charge with respect to the antenna (A) for $n\beta <1$
(not to scale).
              }
         \label{geom1}
   \end{figure}

In this case,  only one instant in the charge's past history has 
a light cone that reaches a given location in space-time
($\BF{r},t$), so that  there is only one retarded time $t_{ret}$
for a given ($\BF{r},t$). Let us take the origin of (observer's)
co-ordinates so that the charge moving along the $z$ axis passes
at $z=0$ at $t=0$, and let us put an electric antenna at 
co-ordinate  $z$  along the charge's path, and distance  $d$   in
the perpendicular plane along the $x$ axis (Fig.\ref{geom1}).
This choice of origin allows to keep the time of arrival of the
charge in the results, which will be used in Sections 4 and 6 to
evaluate coherence effects for a source of finite size.

The distance between the  charge's present and past positions is
PP'$=n\beta R$, so that with the notations of  Fig.\ref{geom1},
P'Q$=n\left( \BFS{\beta}\cdot \BF{n}\right) R$,
whence AQ$=\left( 1-n\BFS{\beta}\cdot \BF{n}\right) R$. From the
triangles CPA and QPA we have $d^2+\left( z+vt\right)
^2=$AQ$^2+\left( n\beta R\sin \theta \right) ^2$, so that, since
$\sin \theta =d/R$, the square of the denominator in the bracket
of (\ref{LW}) is
\EQN{7}{1}{}{}{ \RDLAST{ \CELL{\left[ \left( 1-n \BFS{\beta}\cdot
 \BF{n}\right) R\right] _{ret}^2&= &d^2\left( 1-n^2\beta
^2\right) +\left( z+vt\right) ^2\label{ret}}}{1}{}{}{} {}}
Equation (\ref{ret}) is a purely geometrical relation which holds
whatever the sign of $\left( 1-n^2\beta ^2\right) $,   provided
at least one point P' (source's retarded position) does exist.
This is the case  when the right-hand side term of (\ref{ret}) is
positive or zero, a condition which always holds for the
subluminal case ($n\beta <1$), and which defines the 
\v{C}erenkov cone for $n\beta >1$. We shall consider the latter
case in Section 5.

Substituting (\ref{ret}) into (\ref{LW}) yields
\EQN{7}{1}{}{}{ \RDLAST{ \CELL{\Phi ( \BF{r},t)&= &\frac q{4\pi
\epsilon _0n^2d\left( 1-n^2\beta ^2\right) ^{1/2}}\left[ 1+\left(
\frac{t+z/v}\tau \right) ^2\right] ^{-1/2}\label{phit}}}{1}{}{}{}
{}}
 where
\EQN{7}{1}{}{}{ \RDLAST{ \CELL{\tau &= &d\mid 1-n^2\beta ^2\mid
^{1/2}/v\label{tau}}}{1}{}{}{} {}}
Note that we have used an absolute value in (\ref{tau}), although
$1-n^2\beta ^2>0$ for a subluminal charge, in order to be able to
use the same definition of  $\tau $ for a supraluminal charge
(Section  5).

Equations (\ref{phit}) and (\ref{LWA}) yield a pulse in the
potentials,  centred at $t=-z/v$, the time when the charge's
trajectory passes at closest approach to the antenna, and of
half-width about $\tau $. In the limit $n\rightarrow 1$,
(\ref{phit}) reduces to the well-known potential of a point
charge in uniform motion in vacuum (see for example Feynman
\cite{fey64}, Jackson  \cite{jac99}).

It is important to recall that in (\ref{phit}), the relevant
charge and velocity are retarded quantities. Namely, the
potentials only depend on the charge, position and speed of the
particle at the retarded time. This means that the pulse's peak
is produced by the past state of the charge, at the retarded time
$t_{ret}$, when the charge was at distance $z_{ret}-z=d/\tan
\theta $ from the antenna along the direction of motion. Since at
the time of  the peak of potential at the antenna (A), the
charge's position (P) is at C so that  $\tan \theta =d/\left(
n\beta R\right) $, we have $\cos \theta =n\beta $ at this time,
whence
\EQN{7}{1}{}{}{ \RDLAST{ \CELL{z_{ret}-z&= &n\beta d\left(
1-n^2\beta ^2\right) ^{-1/2}\label{zret}}}{1}{}{}{} {}}

The state of the charge at  times different from the retarded
time that produce the pulse is  irrelevant. In particular, the
instant of the peak is that at which  the  charge moving at
$\BF{v}$ along the $z$ axis passes at closest approach to the
antenna,  but the \emph{real\/} charge need not necessarily be
there at this time; it may have disappeared or have changed its
trajectory; the latter effects do not change the potentials, but
produce an additional contribution to the electric field
(\ref{EphiA}) due to the term  $\partial \BF{A}/\partial 
t$ (see  Feynman  \& al. \cite{fey64}).

For cosmic showers, therefore, the relevant charge and speed
producing the field are those corresponding to the development of
the shower at the retarded altitude, given by (\ref{zret}); hence
they do not necessarily correspond to maximum shower development.
For an antenna on the ground and a shower inclined to the
vertical by the angle $\alpha $, we see from Fig.\ref{geom1} and
Eq.(\ref{zret}) that the retarded altitude is $h_{ret}=$ GP'$\cos
\alpha $, i.e.
\EQN{7}{1}{}{}{ \RDLAST{ \CELL{h_{ret}&= &d\left[ n\beta \left(
1-n^2\beta ^2\right) ^{-1/2}\cos \alpha \pm \sin \alpha \right]
\label{hret0}}}{1}{}{}{} {}}
where the signs $+$ and $-$ correspond respectively to the charge
impacting the ground after or before the time of closest
approach. For $\gamma \gg 1$, $n-1\ll 1$, CP'$=z_{ret}-z$ is much
greater than GC, so that (\ref{hret0}) yields approximately
\EQN{7}{1}{}{}{ \RDLAST{ \CELL{h_{ret}&\simeq & d\left(
1-n^2\beta ^2\right) ^{-1/2}\cos \alpha \label{hret}}}{1}{}{}{}
{}}
where we have also omitted the factor $n\beta \simeq 1$. We shall
return to the retarded altitude later.

From (\ref{phit}), we deduce immediately the $x$  component of
the electric field $E_x=-\partial \Phi /\partial d$
\EQN{7}{1}{}{}{ \RDLAST{ \CELL{E_x&= &\frac q{4\pi \epsilon
_0n^2d^2\left( 1-n^2\beta ^2\right) ^{1/2}}\left[ 1+\left(
\frac{t+z/v}\tau \right) ^2\right] ^{-3/2}\label{Ext}}}{1}{}{}{}
{}}

\subsection{Effect of the refractive index}

\begin{figure}
   \centering
 \includegraphics[width=6cm]{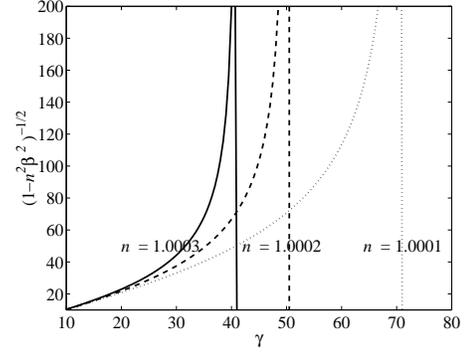}
      \caption{A subluminal charge of Lorentz factor $\gamma $ in
a medium of refractive index $n$ produces a similar electric
potential as a charge in vacuum of Lorentz factor  $\gamma
^\prime $, plotted here versus $\gamma $ for different values of
the air refractive index. When   $\gamma $ approaches the value
for which $n\beta =1$,  $\gamma ^\prime $ increases rapidly and
both the variation of $n$ and the strong increase in retarded
altitude come into play. For still greater values of $\gamma $,  
$\gamma ^\prime $ is  imaginary and \v{C}erenkov radiation is
produced.
           }
       \label{gamma}
   \end{figure}

For relativistic velocities, the refractive index plays an
important role, even though it is close to $1$. Indeed,  we see
from (\ref{phit})-(\ref{tau}) that a subluminal charge $q$ of
Lorentz factor $\gamma =\left( 1-\beta ^2\right) ^{-1/2}$  in a
medium of refractive index $n\neq 1$ produces the same electric
potential as a charge $q/n^2$ of  Lorentz factor 
\EQN{7}{1}{}{}{ \RDLAST{ \CELL{\gamma ^\prime &= &\left(
1-n^2\beta ^2\right) ^{-1/2}\label{gammap}}}{1}{}{}{} {}}
 in vacuum. In particular, the pulse maximum amplitude and
retarded altitude are both proportional to $\gamma ^\prime $,
whereas the time duration is inversely proportional to  $\gamma
^\prime $.

Figure \ref{gamma} shows the factor $\gamma ^\prime $ as a
function of $\gamma $ for different values of the air refractive
index. One sees that, as we noted in the introduction, the
subluminal upper limit $n\beta =1$ takes place at an energy of
the same order of magnitude as the median energy of  shower
charges ($\gamma \simeq 60$). Note that for $\gamma \gg 1$,
$n-1\ll 1$, we have  $1-n^2\beta ^2\simeq 1/\gamma ^2-2\left(
n-1\right) $, so that $\gamma ^\prime $ may be approximated by
\EQN{7}{1}{}{}{ \RDLAST{ \CELL{\gamma ^\prime &\simeq & \gamma
\left[ 1-2\left( n-1\right) \gamma ^2\right]
^{-1/2}\label{gammapp}}}{1}{}{}{} {}}

In practice, since the refractive index decreases as altitude
increases, the retarded distance $R$ should be evaluated by an
integration involving the wave phase speed at different altitudes
from emission to reception. Since the greater the refractive
index, the greater the propagation time, this integral is mainly
determined by the greater values of the index. To simplify the
calculations, we thus approximate the  index by its value at low
to mid altitudes, $n\simeq 1+2\times 10^{-4}$,  but will keep in
mind that $n-1$ is roughly proportional to the air density,
decreasing roughly exponentially with altitude, and increasing
with humidity (see for example Birch \cite{bir94}), which may 
have important consequences.

 With this value of $n$, we see from (\ref{gammapp}) that 
$\gamma ^\prime \simeq \gamma $ only for $\gamma $ significantly
smaller than $\left[ 2\left( n-1\right) \right] ^{-1/2}\simeq
50$, i.e. for electrons of energy significantly smaller than
about 25 MeV.   Since the median electron energy in a typical
shower is about 30 MeV ($\gamma \simeq 60$), the radio electric
field is expected to be rather different from its value in
vacuum. For example, electrons of $\gamma \simeq 35$
(respectively 40) in air with the above value of $n$  produce a
similar electric field as electrons having $\gamma ^\prime \simeq
49$ (respectively 67) in vacuum. 

Close to  $n\beta =1$, i.e. $\gamma \simeq \left[ 2\left(
n-1\right) \right] ^{-1/2}$, $\gamma ^\prime $  varies very
rapidly. In this region, two physical effects come into play:
  \begin{itemize}
 \item the variation of the refractive index, since $\gamma
^\prime $  is very sensitive to $n$ in this region,
 \item the strong increase of the retarded altitude with $\gamma
^\prime $, which tends to put it above the atmosphere, so that
the retarded charge becomes negligible.
\end{itemize}
For example, at perpendicular distance $d=200$ m and vertical
inclination angle $\alpha =\pi /4$, (\ref{hret}) yields
$h_{ret}=$ 20 km for $\gamma ^\prime \simeq 140$, i.e. $\gamma
\simeq 47$. Hence in that case, for the retarded altitude not to
be above the region of shower development, the energy of the
radiating electrons should be smaller than about 24 MeV. This
means that, since the median energy is about 30 MeV), typically a
little less than the lower half of the electron energy
distribution in the shower contribute to the boosted Coulomb
electric field, for these values of perpendicular distance and
angle to the vertical.

For greater energies, the retarded altitude is above the
atmosphere, except for showers of large inclinations or passing
very close to the antenna. And at energies so that $n\beta >1$, 
$\gamma ^\prime $ becomes imaginary; in that case, there is no
longer a single retarded time  and position, and a \v{C}erenkov 
field is produced, which we shall evaluate in Section
\ref{cerenkov}.

\subsection{Electric field spectrum}

Defining the Fourier transforms of the potentials as
\EQN{7}{1}{}{}{ \RDLAST{ \CELL{\Phi ( \BF{r},\omega )&=
&\int_{-\infty }^{+\infty }dt\,e^{i\omega t}\,\Phi \left( 
\BF{r},t\right) \label{TFphi}}}{1}{}{}{} {}}
we have from (\ref{phit}) and (\ref{LWA})
\EQN{7}{1}{}{}{ \RD{ \CELL{\Phi ( \BF{r},\omega )&=
&\frac{qe^{i\omega z/v}}{2\pi \epsilon _0n^2v}K_0(\omega d/\gamma
^\prime v)\label{phiom}}}{1}{}{}{} \RDLAST{ \CELL{ \BF{A}(
\BF{r},\omega )  &= & n^2 \BFS{\beta}\Phi ( \BF{r},\omega
)/c\label{Aom}}}{1}{}{}{}}
 Here $K_0$ is  a modified Bessel function of order 1 (Abramowitz
\& Stegun \cite{abr72}), $\tau $ is given by (\ref{tau}), $z$  is
the antenna's co-ordinate along the charge's path (whose origin
is the charge's co-ordinate at $t=0$), and $d$ is the antenna's
perpendicular distance to the charge's path.

The electric field is given by
\EQN{7}{1}{}{}{ \RDLAST{ \CELL{ \BF{E}\left(  \BF{r},t\right) &=
&-\bigtriangledown \Phi - \BFS{\partial} \BF{A}/\partial
t\label{EphiA}}}{1}{}{}{} {}}
 so that its Fourier transform has from (\ref{phiom})-(\ref{Aom})
the components
\EQN{7}{1}{}{}{ \RD{ \CELL{E_x( \BF{r},\omega )&=
&\frac{q\omega e^{i\omega z/v}}{2\pi \epsilon _0n^2v^2\gamma ^\prime
}K_1(\omega d/\gamma ^\prime
v)\label{Ex}}}{1}{}{}{} \RDLAST{ \CELL{E_z( \BF{r},\omega )  &= &
\frac{-iq\omega e^{i\omega z/v}}{2\pi \epsilon _0n^2v^2 \gamma ^{\prime 2}} K_0(\omega d/\gamma ^\prime
v)\label{Ez}}}{1}{}{}{}}
Therefore $E_z/E_x\ll 1$ for $\gamma \gg 1$, $n-1\ll 1$, so that
 the electric field is radial (perpendicular to $\BF{v}$,
directed along the charge's present position to the antenna) of
amplitude
 \EQN{7}{1}{}{}{ \RDLAST{ \CELL{E( \BF{r},\omega )&\simeq &
\frac{q\omega e^{i\omega z/v}}{2\pi \epsilon _0n^2v^2\gamma
^\prime }K_1\left( \omega d/\gamma ^\prime v\right)
\label{E}}}{1}{}{}{} {}} 
with $\gamma ^\prime $ given by (\ref{gammap}), at perpendicular
distance $d$ from the charge's path.

Using the expansions of the Bessel function $K_1$ (Abramowitz \&
Stegun \cite{abr72}), (\ref{E}) yields at respectively low and
high frequencies (or distances)
\EQN{7}{1}{}{}{ \RD{ \CELL{\frac{\omega d}{\gamma ^\prime v}&\ll
& 1\,\,\,\,\,E( \BF{r},\omega ) \simeq   \frac{qe^{i\omega
z/v}}{2\pi \epsilon _0n^2vd}\label{Elf}}}{1}{}{}{} \RDLAST{
\CELL{\frac{\omega d}{\gamma ^\prime v}&\gg & 1\,\,\,\,E(
\BF{r},\omega ) \simeq   \frac{qe^{i\omega z/v}}{4\pi \epsilon
_0n^2v^2}\left( \frac{2\pi \omega v}{\gamma ^\prime d}\right)
^{1/2}e^{-\omega d/\gamma ^\prime v}\label{Ehf}}}{1}{}{}{}}
 for $n-1\ll 1$, $\beta \simeq 1$.  The low-frequency (or small
distance) spectrum (the time integral of the pulse) is
independent of  $\gamma ^\prime $, as expected since the larger
$\gamma ^\prime $, the greater the amplitude of the pulse, but
the smaller (in the same proportion) the duration. At large
frequencies and/or distances, the field decreases nearly
exponentially with the product  $\omega d/\gamma ^\prime v$, thus
with a frequency scale proportional to $1/d$ and a distance scale
proportional to $1/\omega $. At intermediate frequencies, the
factor $\sqrt{\omega /d}$ in (\ref{Ehf}) makes the decrease with $\omega $
(respectively $d$) slower (respectively faster) than given by the
exponential  $e^{-\omega d/\gamma ^\prime v}$ alone. 

\begin{figure}
   \centering
 \includegraphics[width=7cm]{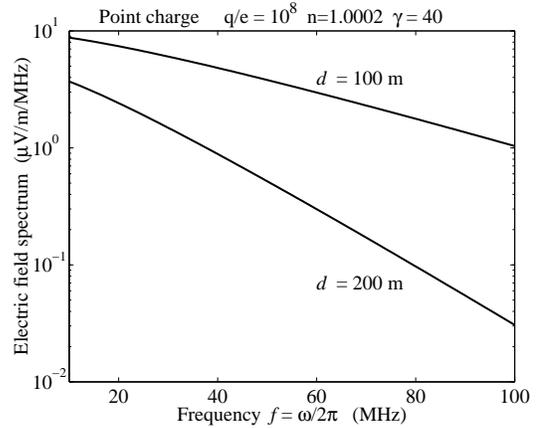}
      \caption{Spectral field strength $\mid E(\omega )\mid $
from (\ref{E}), produced by a point charge of $10^8$ electrons at
perpendicular distances $d=100$ and 200 m, for $n=1.0002$ and
$\gamma =40$. 
             }
         \label{Epointom}
   \end{figure}

Figure \ref{Epointom} shows the modulus of the electric field
spectrum (\ref{E}) produced by a charge $q=Ne$ with $N=10^8$ ($e$
being  the electron charge) for $n=1.0002$ and  $\gamma =40$
($\gamma ^\prime \simeq 67$), at perpendicular distances $d=100$
and 200 m, for which the retarded altitude (\ref{hret}) is
respectively 5 and 10 km at a vertical angle $\alpha =\pi /4$.

Note that the plotted values correspond to Fourier transforms
(defined as in (\ref{TFphi})), so that they should be multiplied
by a factor of two when comparing them to measured spectra, which
are generally defined for positive frequencies only.

\section{Charge at the retarded altitude\label{altret}}

The retarded charge producing the pulse is that corresponding to
the shower development at the retarded altitude, given by
(\ref{hret}) as  a function of the perpendicular distance $d$ to
the path and of the inclination $\alpha $ to the vertical. In the
region of maximum shower development,  the number of charged
particles (mainly electrons and positrons) is
\EQN{7}{1}{}{}{ \RDLAST{ \CELL{N_M&\simeq &0.31\left(
W_p/W_c\right)
\left[ \ln \left( W_p/W_c\right) \right]
^{-1/2}\label{NM}}}{1}{}{}{} {}}
for a primary of energy $W_p$, where $W_c=0.86\times 10^8$ eV
(see for example Abu-Zayyad, \& al. \cite{abu01} and references
therein).

With a simplified atmosphere model of density $\rho
=\rho_0e^{-h/H}$ at altitude $h$, the total mass from the top of
the atmosphere down to altitude $h$ is
\EQN{7}{1}{}{}{ \RDLAST{ \CELL{X(h)&= &\left( \rho _0H/\cos
\alpha \right) e^{-h/H}\label{xh}}}{1}{}{}{} {}}
where $H\simeq 8500$ m and $\rho _0\simeq 1.22$ kg/m$^3$. The
altitude $h_M$ where the shower charge is maximum is determined
by $X(h_M)=X_M$,  where $X_M\simeq 6500$ kg/m$^2$ for a primary
cosmic ray of energy $10^{17}-10^{18}$ eV, and increases at
higher energies (Abu-Zayyad, \& al. \cite{abu01} and references
therein). The number of charged particles at the retarded
altitude is determined by the value of $X$. It  can be obtained
from the Greisen function as
\EQN{7}{1}{}{}{ \RDLAST{ \CELL{N(X)/N_M&= &\exp \left[ \frac
X{L_C}\left( 1-\frac 32\ln s\right) -\frac{X_M}{L_C}\right]
\label{greisen}}}{1}{}{}{} {}}
where  the so-called shower age is $s=3/\left( 1+2X_M/X\right) $,
$N_M$ is given by (\ref{NM}), and  $L_C\simeq 587$
kg/m$^2$ ( Abu-Zayyad, \& al. \cite{abu01}).

\begin{figure}
   \centering
 \includegraphics[width=7cm]{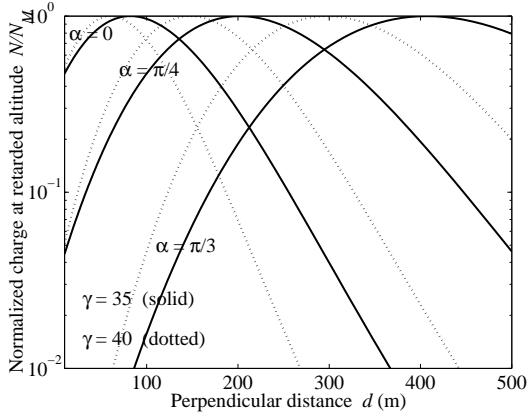}
      \caption{Number of  secondary charges in the shower
(normalised to the value at maximum shower development) at the
retarded altitude  $\gamma ^\prime d\cos \alpha $ for $\gamma
=35$ (solid lines) and  $\gamma =40$ (dotted lines) with
$n=1.0002$ (i.e. respectively $\gamma ^\prime =49$ and 67) as a
function of  distance for different angles $\alpha $ to the
vertical, for a primary of  $10^{17}-10^{18}$ eV.
             }
         \label{charge}
   \end{figure}

Figure  \ref{charge} shows $N/N_M$ as a function of the
perpendicular distance $d$ for different values of the
inclination $\alpha $ of the shower to the vertical, for $\gamma
=35$ and 40 with  $n-1=2\times 10^{-4}$ (respectively $\gamma
^\prime =49$ and 67). 

At normal incidence, the number of charges at the  retarded 
altitude is of the same order of magnitude as the value at
maximum development, up to distances of a few hundred metres, for
which the retarded altitude goes above the region of significant
shower  development. As the angle $\alpha $  increases, the
distance $d$ for which the retarded altitude is in the region of
maximum development increases. The larger $\gamma $ and $\alpha
$, the larger the value of $d$  for which the retarded altitude
corresponds to maximum shower development. As $\gamma $ increases
significantly above 45 (for $n-1\simeq 2\times 10^{-4}$), the
fast increase of $\gamma ^\prime $ puts the retarded altitude
above the atmosphere.

In the following Section, we will thus consider charges of
Lorentz factor  $\gamma =40$. 

\section{Boosted Coulomb spectrum of a charge
distribution\label{finite}}

  \begin{figure}
   \centering
 \includegraphics[width=5cm]{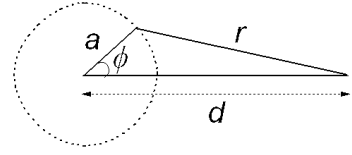}
      \caption{Geometry for calculating the field at
perpendicular distance $d$ from the centre of a charge
distribution with cylindrical symmetry around the charge path.
              }
         \label{geomcharge}
   \end{figure}

In the previous Sections, we have calculated the electric field
of a point charge. However, the charge distribution in the shower
is believed to be pancake shaped, of axis the shower direction,
and of lateral size increasing with altitude. The half-size
containing the bulk of the particles  is about 20-30 m at ground
level, somewhat smaller than the canonical Moli\`{e}re radius of
about 80 m (Antoni  \& al. \cite{ant01}), and  very thin in the
longitudinal direction, with a dispersion in particle arrival
times of about 1.6 ns at the centre (Linsley \cite{lin86}). This
small longitudinal dispersion will decrease the field level
because of the loss of coherence due to the  factor $e^{i\omega
z/v}$ in the spectrum (\ref{E}). 

With an axial symmetry, it is convenient to consider first a
charge distributed round a circle of radius $a$ in the plane
perpendicular to the path, infinitely thin in the $z$  direction,
which allows an analytical calculation. The field of an arbitrary
charge distribution can then be obtained by performing an
integration over the lateral and longitudinal dimensions.

\subsection{Ring charge distribution\label{ringsec}}

Consider a charge $q$  distributed round  a circle of radius $a$
in the plane perpendicular to $z$  which passes at $z=0$  at
$t=0$. The potential produced at perpendicular distance $d$ from
the centre of the circle (and co-ordinate $z$ along the charge's
path) is, from (\ref{phiom}) with the geometry defined in
Fig.\ref{geomcharge}
\EQN{7}{1}{}{}{ \RD{ \CELL{\Phi _a( \BF{r},\omega )&=
&\frac{qe^{i\omega z/v}}{2\pi \epsilon
_0n^2v}F(a)\label{phia}}}{1}{}{}{} \RD{ \CELL{F\left( a\right)  
&= & \frac 1{2\pi }\int_0^{2\pi }d\phi \,K_0(\omega r/\gamma
^\prime v)\label{F}}}{1}{}{}{} \RDLAST{ \CELL{r  &= & \left(
a^2+d^2-2ad\cos \phi \right) ^{1/2}}}{1}{}{}{}}
The integral $F\left( a\right) $  can be calculated analytically
by expanding $K_0$ in series of Bessel functions using Neumann's
addition theorem (Watson  \cite{wat66}), from which we derive
finally
\EQN{7}{1}{}{}{ \RD{ \CELL{F\left( a\right) &= &I_0(\omega
d/\gamma ^\prime v)\,K_0(\omega a/\gamma ^\prime v)\,\,\,\,\,d < 
a\label{F1}}}{1}{}{}{} \RDLAST{ \CELL{F\left( a\right) &=
&I_0(\omega a/\gamma ^\prime v)\,K_0(\omega d/\gamma ^\prime
v)\,\,\,\,\,d >  a\label{F2}}}{1}{}{}{}}
 We deduce from (\ref{EphiA})  the (radial) electric field
\EQN{7}{1}{}{}{ \RD{ \CELL{E( \BF{r},\omega )&= &\frac{-q\omega
e^{i\omega z/v}}{2\pi \epsilon _0n^2v^2\gamma ^\prime }K_0(\omega
a/\gamma ^\prime v)\,I_1(\omega d/\gamma ^\prime v)\,\,\,\,\,d < 
a\label{E1}}}{1}{}{}{} \RDLAST{ \CELL{E( \BF{r},\omega )&=
&\frac{q\omega e^{i\omega z/v}}{2\pi \epsilon _0n^2v^2\gamma
^\prime }I_0(\omega a/\gamma ^\prime v)\,K_1(\omega d/\gamma
^\prime v)\,\,\,\,\,d >  a\label{E2}}}{1}{}{}{}}
which has a discontinuity on the circle (as expected), of no
physical consequence since it  disappears on integration over the
radius $a$. Comparing to the electric field spectrum (\ref{E}) of
a point charge, we see that the finite size of the source removes
the singularity of the field at $d\rightarrow 0$. It also 
increases the field strength outside the source by the factor
$I_0(\omega a/\gamma ^\prime c)$, which increases as frequency
increases, so that the lateral size of the charge makes the
spectrum at large distances fall less quickly with frequency than
that of a point charge.

\subsection{Pancake charge distribution\label{pansec}}

Consider now a charge $q$ of lateral distribution $\sigma (a)$
and finite longitudinal thickness. The potential at perpendicular
distance $d$  from the centre is obtained by integrating 
(\ref{phia}) as
\EQN{7}{1}{}{}{ \RDLAST{ \CELL{\Phi ( \BF{r},\omega )&=
&\int_0^\infty da\,2\pi a\,\sigma (a)\,G\left( a\right) \mid \Phi
_a( \BF{r},\omega )\mid }}{1}{}{}{} {}}
for a distribution normalised so that $2\pi \int_0^\infty
da\,a\,\sigma (a)=1$. Here $G\left( a\right) $ involves an
integral of   the phase factor $e^{i\omega z/v}$ in
Eq.(\ref{phia}), ponderated by the longitudinal source
distribution. The longitudinal extension therefore decreases the
spectrum at frequencies $\omega >1/\delta t$, where  $\delta t$
is the width of the particle arrival time distribution, because
of loss of coherence, so that we have $G\simeq 1$ for $\omega
\delta t<1$, whereas $G\ll 1$ for $\omega \delta t\gg 1$ when the
fields no longer add in phase.

\begin{figure}
   \centering
 \includegraphics[width=7cm]{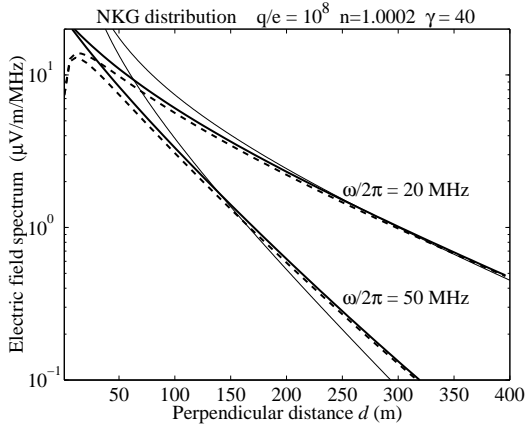}
      \caption{Spectral field strength $\mid E(\omega )\mid $ at 
$\omega /2\pi =20$  and 50 MHz produced by $10^8$ electrons with
a NKG distribution with $s=1$ (respectively 1.4) and $r_M=100$ m
(respectively 50 m), in solid lines (respectively dashed), for
$n=1.0002$ and $\gamma =40$, as a function of perpendicular
distance $d$ to the charge's path, compared to the field of a
point charge (thin lines).
              }
         \label{NKG}
   \end{figure}

We model the lateral distribution $\sigma (a)$ with the so-called
NKG lateral density parametrization $\sigma (a)\propto
(a/r_M)^{s-2}\left( 1+a/r_M\right) ^{s-4.5}$ (see Nishimura
\cite{nis67}). Here the two parameters $s$  and $r_M$ are not
necessarily the conventional shower age and Moli\`{e}re radius
respectively, since a wide set of couples $\left[ s,r_M\right] $
can fit the observed lateral distribution (see Antoni  \& al.
\cite{ant01} and refs. therein). However, the coupled variation
of these parameters keeps roughly the same proportion of
particles within a given lateral distance; for example, KASCADE
measurements shown in Fig. 4 of the paper cited above find nearly
 50\% (respectively 80\%) of the electrons within 20 m
(respectively 50 m) lateral distance.  

The distribution of arrival times is still less well known.
Assuming $\delta t\sim 1.6$ ns at the centre of the pancake, and
a value roughly three times greater at 30 m lateral distance
(Linsley \cite{lin86}), the factor  $G$ should decrease below
unity  at frequencies above about 50 MHz. In this highly
simplified calculation, we will therefore neglect this effect and
the curvature of the pancake, keeping in mind that the
corresponding loss of coherence should decrease the spectrum
above about 50 MHz.

Figure \ref{NKG} shows the boosted Coulomb electric field
spectrum produced by $10^8$ electrons with a NKG distribution
with two sets of parameters: [$s=1,r_M=100$ m]  and 
[$s=1.4,r_M=50$ m], at frequencies  $\omega /2\pi =20$ and 50
MHz, as a function of the perpendicular distance $d$ to the
charge's path, for $n=1.0002$ and $\gamma =40$,  compared to the
field of a point charge. One sees that the degeneracy of the
parameters of the lateral distribution only affects the field at
very short radial distances, where it is smaller than that of a
point charge by a factor depending on the parameters of the
distribution. At distances greater than about 50 m, the field
does not depend significantly on the details of the charge
distribution, in the frame of this highly simplified model.

\subsection{Charge separation by the Lorentz
force\label{Lorentz}}

If there were no systematic separation of positive and negative
charges by the Earth's magnetic field, the net charge producing
the radio pulse would be equal to $q\simeq -eN(X)\eta /2$, where
$N(X)$ is given in (\ref{greisen}) and $\eta $  is the relative
excess of electrons over positrons in the shower, which amounts
to about 20 \% at low energies.

However, the Lorentz force due to the Earth's magnetic field
$\BFS{B}$ separates the positive and negative charges in the
direction $\BFS{v}\times  \BFS{B}$. The boosted Coulomb  electric
field is thus the geometric sum of those produced by the positive
and negative charges, each being directed along the perpendicular
distance to their respective charge's barycentre, with opposite
signs. The net electric field is then no longer directed along
the perpendicular distance to the centre of the shower. For
example, for a vertical shower impacting North of an antenna,
this produces a Coulomb field having a East-West component,  of
strength roughly proportional to the separation between positive
and negative charges.

Let us estimate the  average displacement $L$ for  electrons of
Lorentz factor $\gamma $. The force  $\BFS{F} =- e\BFS{v} \times 
\BFS{B} $ acting during the time $t$  produces a displacement
$Ft^2/(2\gamma m_e)$, in the plane perpendicular to $\BFS{v}$,
directed normal to the projection of  $\BFS{B}$ in this plane.
The time for travelling a radiation length $L_0$ divided by the air density $\rho $ is $\Delta t = L_0/(\rho c)$, and the average of $t^2$ is  $2\Delta t^2$, so
that  the average displacement of electrons is 
\EQN{7}{1}{}{}{ \RDLAST{ \CELL{L&\simeq & \frac{eBL_0^2\sin \beta
_0}{\gamma m_ec\rho ^2}}}{1}{}{}{} {}}
where $\beta _0$ is the angle between the  magnetic field and the
shower direction. With $L_0\simeq 367$ 
kg/m$^2$, so that $L_0/(\rho_0 c) \simeq 10^{-6} $ s at sea level, and $B\simeq 0.4$ G, this yields
\EQN{7}{1}{}{}{ \RDLAST{ \CELL{L&\simeq & \frac{2100\sin \beta
_0}{\gamma \left( \rho /\rho _0\right)
^2}\,\,\text{m}\label{L}}}{1}{}{}{} {}}
With $\beta _0=\pi /4$  and $\gamma =40$, this yields  $L\simeq
40$ m at sea level and  $L\simeq 60$ m at an altitude of half the
atmospheric scale height. Positrons are displaced in the opposite
direction, producing a separation of $2L$. Allan  (\cite{all71})
finds values somewhat greater, but of the same order of
magnitude. 

\begin{figure}
   \centering
 \includegraphics[width=7cm]{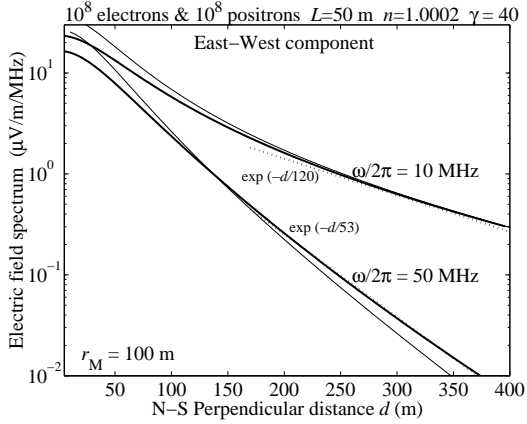}
      \caption{Spectral field strength projected on the East-West
direction,  produced at  $\omega /2\pi =10$  and 50 MHz by 
$10^8$ electrons and  $10^8$ positrons  in vertical motion, 
displaced laterally by 50 m in opposite senses along the
East-West direction, with  NKG distributions with $s=1$  and
$r_M=100$ m (solid lines) compared to point charges (thin lines),
for $n=1.0002$ and $\gamma =40$,  as a function of distance $d$ 
to the centre along the NS axis.  Exponential decreases are shown
for comparison (dotted).          }
         \label{NKGdEWcf}
   \end{figure}

As an example, Fig. \ref{NKGdEWcf} shows the electric field
spectrum projected on the East-West direction, as a function of
distance $d$  to the centre along the NS axis, produced at 
$\omega /2\pi =10$  and 50 MHz by  $10^8$ electrons and  $10^8$
positrons in vertical motion, displaced laterally by 50 m in
opposite senses along the East-West direction,  for $n=1.0002$
and $\gamma =40$ ($\gamma ^\prime =67$); each charge species is
assumed to have a NKG distribution with $s=1$  and $r_M=100$ m
(solid lines). Exponential decreases are shown for comparison
(dotted). Comparing to the case of point charges (thin lines),
one sees that in this simple model, the field is not much
affected by the NKG lateral distribution, except at very short
distances. 

\begin{figure}
   \centering
 \includegraphics[width=9cm]{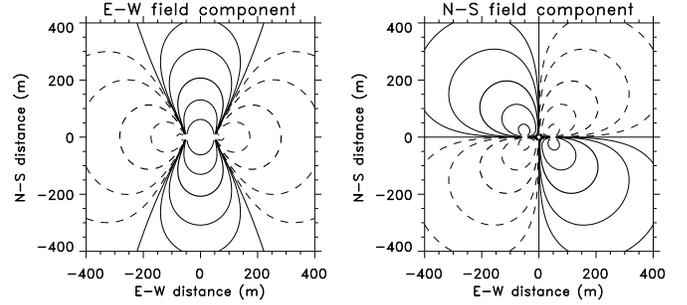}
      \caption{Contour plots of the East-West and North-South
components of the boosted Coulomb electric field spectrum,
produced at $\omega /2\pi =$ 20 MHz by  $10^8$ electrons and 
$10^8$ positrons in vertical motion, displaced laterally by 50 m
in opposite senses along the Est-West direction, for $n=1.0002$
and $\gamma =40$. Contours levels correspond to strengths of  0,
0.1, 0.3, 1 (thicker lines), 3, 10 $\mu $V/m/MHz, with solid and
dashed lines corresponding to components of different signs.
           }
         \label{NSEW}
   \end{figure}

As an example of the change in polarisation introduced by the
magnetic field, Fig. \ref{NSEW} shows contours of   the 
East-West and North-South  components of the boosted Coulomb
electric field spectrum produced at 20 MHz by $10^8$ electrons
and  $10^8$ positrons  in vertical motion,  displaced laterally
by 50 m in opposite senses along the East-West direction;   solid
and dashed lines distinguish the sign of the field components;
$\gamma $ and $n$ have the same values as in Fig.\ref{NKGdEWcf}.

Note that the excess of electrons over positrons, not taken into
account in this plot, produces an additional contribution
changing both polarisation and amplitude.

\section{Electric field spectrum of a supraluminal
charge\label{cerenkov}}

Consider now a point charge moving faster than the waves, so that
 $n\beta >1$. This will produce a \v{C}erenkov field instead of a
boosted Coulomb field.

\subsection{Retarded potentials}

As we noted in Section 2, the retarded potentials are still given
by Eqs.(\ref{LW})-(\ref{ret}), but the geometry is different from
Fig. \ref{geom1}  since now PP' $=n\beta R>R$. This means that P
is on the other side of C on the $z$  axis, at a distance so that
$\cos \theta \geq 1/n\beta $. Therefore, the retarded source's 
distance $R$   is now given by
\EQN{7}{1}{}{}{ \RDLAST{ \CELL{R^2&= &d^2+\left[ n\beta R-\left(
z+vt\right) \right] ^2}}{1}{}{}{} {}}
which has  two solutions if
\EQN{7}{1}{}{}{ \RDLAST{ \CELL{z+vt&> &d\left( n^2\beta
^2-1\right) ^{1/2}}}{1}{}{}{} {}}
i.e. $t+z/v>\tau $  with $\tau $ given by  (\ref{tau}). When this
inequality holds, the antenna is inside  the \v{C}erenkov cone of
half-angle $\arcsin \left( 1/n\beta \right) $ trailing the
charge. When the opposite inequality holds, there is no solution
for the retarded source, whereas when the two terms are equal
(antenna on the \v{C}erenkov cone), the two solutions merge
together.

This means that as time increases, the antenna first sees nothing
(when it is still outside the \v{C}erenkov cone), then it sees
the field of one retarded source (when it is on the  \v{C}erenkov
cone), which splits in two as the  \v{C}erenkov cone trailing the
charge moves so that the antenna is inside it.

Substituting (\ref{ret}) into (\ref{LW}) and summing on the
retarded sources yields the  potential
\EQN{7}{1}{}{}{ \RDLAST{ \CELL{\Phi ( \BF{r},t)&= &\frac q{2\pi
\epsilon _0n^2d\left( n^2\beta ^2-1\right) ^{1/2}}\left[ \left(
\frac{t+z/v}\tau \right) ^2-1\right]
^{-1/2}\label{phitc}}}{1}{}{}{} {}}
for 
\EQN{7}{1}{}{}{ \RDLAST{ \CELL{t+z/v&> &\tau
\label{ttau}}}{1}{}{}{} {}}
and  $\Phi (\BF{r},t)=0$  otherwise, which is similar to a result
by Jackson (\cite{jac99}). The potential is singular at
$t+z/v=\tau $, when the antenna is on the \v{C}erenkov cone of
the charge; at this time there is only one retarded source's
position, satisfying $z_{ret}-z=d/\tan \theta $ with $\cos \theta
=1/n\beta $, i.e
\EQN{7}{1}{}{}{ \RDLAST{ \CELL{z_{ret}-z&= &d\left( n^2\beta
^2-1\right) ^{-1/2}\label{zretc}}}{1}{}{}{} {}}
so that the retarded altitude is now approximately
\EQN{7}{1}{}{}{ \RDLAST{ \CELL{h_{ret}&\simeq & d\left( n^2\beta
^2-1\right) ^{-1/2}\cos \alpha \label{hretc}}}{1}{}{}{} {}}
for a shower of vertical inclination $\alpha $ with $\gamma \gg
1$, $n-1\ll 1$.

\begin{figure}
   \centering
 \includegraphics[width=6cm]{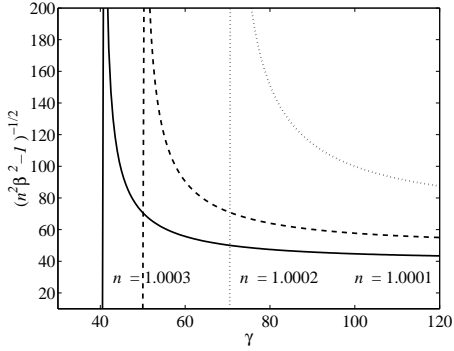}
      \caption{Values of $\gamma _c=\left( n^2\beta ^2-1\right)
^{-1/2}$ as a function of $\gamma =\left( 1-\beta ^2\right)
^{-1/2}$ for different values of the refractive index under
supraluminal conditions.        }
         \label{gammac}
   \end{figure}

We have plotted on Fig. \ref{gammac}
\EQN{7}{1}{}{}{ \RDLAST{ \CELL{\gamma _c&= &\left( n^2\beta
^2-1\right) ^{-1/2}\label{gammacc}}}{1}{}{}{} {}}
as a function of $\gamma $. For $n-1=2\times 10^{-4}$, particles
with $\gamma >50$ produce a \v{C}erenkov field; however, one sees
from Fig. \ref{charge} that, for the retarded altitude
$h_{ret}\simeq \gamma _cd\cos \alpha $ to ensure a significant
retarded charge when $d\simeq 100-200$ m, $\gamma _c$ should not
be smaller than about 70; from Fig. \ref{gammac} this requires
$\gamma >71$ with the above value of $n$.

This means that a little less than the higher half of the energy
distribution can contribute to the \v{C}erenkov field: the
particles for which $n\beta >1$ and whose retarded altitude is in
the region of significant shower development. However, since the
excess of electrons over positrons and the charge separation by
the magnetic field (\ref{L}) both decrease with energy,   only a
small proportion of  these charges should contribute to the
\v{C}erenkov field.

From (\ref{phitc}), we deduce immediately the $x$  component of
the electric field $E_x=-\partial \Phi /\partial d$
\EQN{7}{1}{}{}{ \RDLAST{ \CELL{E_x&= &\frac{-q}{2\pi \epsilon
_0n^2d^2\left( n^2\beta ^2-1\right) ^{1/2}}\left[ \left(
\frac{t+z/v}\tau \right) ^2-1\right]
^{-3/2}\label{Extc}}}{1}{}{}{} {}}
when (\ref{ttau}) holds and 0 when the opposite inequality holds.

\subsection{\v{C}erenkov electric field spectrum}

Calculating the Fourier transform of the potential, which is
given by (\ref{phitc}) when the inequality (\ref{ttau}) holds and
vanishes otherwise, yields the potentials in Fourier space
\EQN{7}{1}{}{}{ \RD{ \CELL{\Phi ( \BF{r},\omega )&=
&\frac{iqe^{i\omega z/v}}{4\epsilon _0n^2v}H_0^1(\omega d/\gamma
_cv)\label{phiomc}}}{1}{}{}{} \RDLAST{ \CELL{ \BF{A}(
\BF{r},\omega )  &= & n^2 \BFS{\beta}\Phi ( \BF{r},\omega
)/c\label{Aomc}}}{1}{}{}{}}
 Here $H_0^1=J_0+iY_0$ is  a Hankel function of the third kind
and order 1(Abramowitz \& Stegun \cite{abr72}), $z$ is the
antenna's co-ordinate along the charge's path (whose origin is
the charge's co-ordinate at $t=0$), and $d$ is the antenna's
perpendicular distance to the charge's path.

The electric field is given by substituting the potentials 
(\ref{phiomc})-(\ref{Aomc}) into (\ref{EphiA}), so that its
Fourier transform has the components
\EQN{7}{1}{}{}{ \RD{ \CELL{E_x( \BF{r},\omega )&= &\frac{iq\omega
e^{i\omega z/v}}{4\epsilon _0n^2v^2\gamma _c}H_1^1(\omega
d/\gamma _cv)\label{Exc}}}{1}{}{}{} \RDLAST{ \CELL{E_z(
\BF{r},\omega )  &= & \frac{-q\omega e^{i\omega z/v}}{4\epsilon
_0n^2v^2\gamma _c^2}H_0^1(\omega d/\gamma
_cv)\label{Ezc}}}{1}{}{}{}}
where $H_1^1=J_1+iY_1$.  Therefore $E_z/E_x\ll 1$ for $\gamma
_c\gg 1$, $n-1\ll 1$, so that  the electric field is nearly 
radial (from the charge's present position to the antenna), of
amplitude
 \EQN{7}{1}{}{}{ \RDLAST{ \CELL{E( \BF{r},\omega )&\simeq &
\frac{iq\omega e^{i\omega z/v}}{4\epsilon _0n^2v^2\gamma
_c}H_1^1(\omega d/\gamma _cv)\label{Ec}}}{1}{}{}{} {}} 
with $\gamma _c$ given by (\ref{gammacc}).

Using the expansions of the Bessel function $H_1^1$ (Abramowitz
\& Stegun \cite{abr72}), Eq.(\ref{Ec}) yields at respectively low
and high frequencies (or distances)\EQN{7}{1}{}{}{ \RD{
\CELL{\frac{\omega d}{\gamma _cv}&\ll & 1\,\,\,E( \BF{r},\omega )
\simeq   \frac{qe^{i\omega z/v}}{2\pi \epsilon
_0n^2vd}\label{Elfc}}}{1}{}{}{} \RDLAST{ \CELL{\frac{\omega
d}{\gamma _cv}&\gg & 1\,\,\,\,E( \BF{r},\omega ) \simeq  
\frac{qe^{i(\omega z/v-\pi /4)}}{4\pi \epsilon _0n^2v^2}\left(
\frac{2\pi \omega v}{\gamma _cd}\right) ^{1/2}e^{i\omega d/\gamma
_cv}\nonumber}}{1}{}{}{}}
for $\gamma _c\gg 1$, $n-1\ll 1$.

In the low-frequency (or small distance) limit, the electric
field has the same expression as for $v<c/n$, as expected since
it is independent of $\gamma $  in this limit. This was already
noted by Allan (\cite{all71}). On the other hand, at large
frequencies (or distances), the exponential decrease is replaced
by a phase oscillation, again as expected since $\gamma ^\prime $
is replaced by $i\gamma _c$ when $v>c/n$. This phase oscillation
suggests that the finite lateral size of the source should
decrease the field with respect to that of a point charge,
instead of increasing it as occurred for the boosted Coulomb
field, because of loss of coherence. This effect is estimated in
the next Section. Furthermore, the large frequency (or distance)
value is essentially produced by the fast variation (see
Eq.(\ref{Extc})) near $t+z/v=\tau $ when the antenna is on the
\v{C}erenkov cone; this contribution is expected to be sensitive
to the variation of the refractive index. 

\section{\v{C}erenkov spectrum of a charge distribution}

Consider a charge $q$  distributed round  a circle of radius $a$
in the plane perpendicular to $z$  which passes at $z=0$  at
$t=0$. The potential produced at perpendicular distance $d$ from
the centre of the circle (and distance $z$ along the charge's
path) is, from (\ref{phiomc}) with the geometry defined in
Fig.\ref{geomcharge}
\EQN{7}{1}{}{}{ \RD{ \CELL{\Phi _a( \BF{r},\omega )&=
&\frac{iqe^{i\omega z/v}}{4\epsilon
_0n^2v}F_c(a)\label{phiac}}}{1}{}{}{} \RD{ \CELL{F_c\left(
a\right)   &= & \frac 1{2\pi }\int_0^{2\pi }d\phi \,H_0^1(\omega
r/\gamma _cv)\label{Fc}}}{1}{}{}{} \RDLAST{ \CELL{r  &= & \left(
a^2+d^2-2ad\cos \phi \right) ^{1/2}}}{1}{}{}{}}
 The integral $F_c\left( a\right) $  can be calculated
analytically by expanding $H_0^1$ in series of Bessel functions
using Neumann's addition theorem (Watson  \cite{wat66}), from
which we derive finally
\EQN{7}{1}{}{}{ \RD{ \CELL{F_c\left( a\right) &= &J_0(\omega
d/\gamma _cv)\,H_0^1(\omega a/\gamma _cv)\,\,\,\,\,d < 
a\label{F1c}}}{1}{}{}{} \RDLAST{ \CELL{F_c\left( a\right) &=
&J_0(\omega a/\gamma _cv)\,H_0^1(\omega d/\gamma _cv)\,\,\,\,\,d
>  a\label{F2c}}}{1}{}{}{}}
 
We deduce from (\ref{EphiA})  the electric field along the
perpendicular distance
\EQN{7}{1}{}{}{ \RD{ \CELL{E_x( \BF{r},\omega )&= &\frac{iq\omega
e^{i\omega z/v}}{4\epsilon _0n^2v^2\gamma _c}H_0^1(\omega
a/\gamma _cv)J_1(\omega d/\gamma _cv)\,\,\,\,\,d < 
a\label{E1c}}}{1}{}{}{} \RDLAST{ \CELL{E_x( \BF{r},\omega )&=
&\frac{iq\omega e^{i\omega z/v}}{4\epsilon _0n^2v^2\gamma
_c}J_0(\omega a/\gamma _cv)H_1^1(\omega d/\gamma _cv)\,\,\,\,\,d
>  a\label{E2c}}}{1}{}{}{}}
and along the charge's path
\EQN{7}{1}{}{}{ \RD{ \CELL{E_z( \BF{r},\omega )&= &\frac{q\omega
e^{i\omega z/v}}{4\epsilon _0n^2v^2}H_0^1(\omega a/\gamma
_cv)J_0(\omega d/\gamma _cv)\,\,\,\,\,d < 
a\label{E1cz}}}{1}{}{}{} \RDLAST{ \CELL{E_z( \BF{r},\omega )&=
&\frac{q\omega e^{i\omega z/v}}{4\epsilon _0n^2v^2}J_0(\omega
a/\gamma _cv)H_0^1(\omega d/\gamma _cv)\,\,\,\,\,d > 
a\label{E2cz}}}{1}{}{}{}}
The expressions (\ref{E1c})-(\ref{E2c}) are similar to those
derived by  Kahn  \& Lerche (\cite{kah66}) with a different
formulation involving the wave equation; note that the latter
study did not include the boosted Coulomb field since the charge
speed was assumed to be equal to that of the primary.

As in the case $v<c/n$, the $x$  component (radial in cylindrical
co-ordinates around the charge's path) is much greater than the
$z$ one when $\gamma _c\gg 1$. 

Comparing to the electric field spectrum (\ref{Ec}) of a point
charge, we see that, as for the boosted Coulomb field, the finite
size of the source removes the singularity at $d\rightarrow 0$,
yielding $E=0$ at the centre (as it should by symmetry). Note
also that, contrary to the Coulomb case, the finite lateral size
of the source decreases the field strength, by the factor
$J_0(\omega a/\gamma _cv)$.

\begin{figure}
   \centering
 \includegraphics[width=6cm]{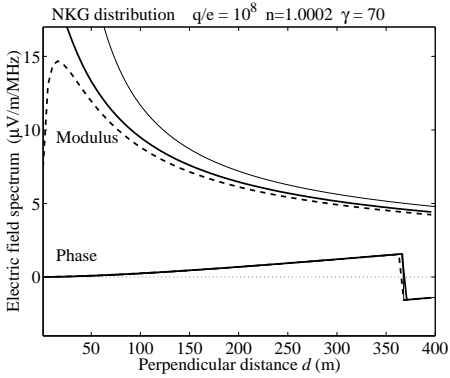}
      \caption{Spectral field produced at frequency  $\omega
/2\pi =$ 20 MHz by $10^8$ electrons with  NKG distributions with
$s=1$  and $r_M=100$ m (solid line) and  $s=1.4$  and $r_M=50$ m
(dashed  line) for $n=1.0002$ and $\gamma =70$ at a function of
perpendicular distance  $d$, compared to the field of a point
charge (thin line).
             }
         \label{NKGc}
   \end{figure}

This is illustrated in Fig. \ref{NKGc} , which shows the modulus
and phase of the \v{C}erenkov electric field produced at
frequency  $\omega /2\pi =$ 20 MHz by $10^8$ electrons with a NKG
distribution with the two same sets of parameters as in
Fig.\ref{NKG}, for $n=1.0002$ and $\gamma =70$ at perpendicular
distance  $d=$ 100 m, compared to the field of a point charge
(thin lines).

Another difference with the Coulomb field is that the phase
varies rapidly with the charge energy, except at small
frequencies and/or distances, since $H_1^1(\zeta )\sim \left(
2/\pi \zeta \right) ^{1/2}e^{i\left( \zeta -3\pi /4\right) }$ for
$\zeta \gg 1$; hence we see from (\ref{E2c}) that integrating
over a distribution of $\gamma $ should decrease the field
strength.

\section{Comparison with the \v{C}erenkov radiation}

We have seen that the electric field is essentially directed
along the vector from the charge's present position to the
antenna ($\BF{x}$ axis), i.e. perpendicular to the velocity.
However, it is not this field that is at the origin of
conventional \v{C}erenkov radiation, but instead the (smaller)
component parallel to  the velocity ($\BF{z}$ axis).

Indeed, the energy flow through a cylinder of radius $d$ around
the charge's path is given by
\EQN{7}{1}{}{}{ \RDLAST{ \CELL{\frac{dW}{dz}&=
&\frac{-1}v\frac{dW}{dt} =  \frac{\epsilon _0c}v\int_{-\infty
}^{+\infty }dz\,2\pi d\,E_z\,B_y}}{1}{}{}{} {}}
 since the wave magnetic field $\BFS{B} = \nabla \times \BFS{A}$ 
with $\BF{A}$ given by (\ref{Aomc}) is along $\BF{y}$.   This
integral, along the cylinder at a given time, may be transformed
into an integral over time at a given point on the cylinder
(using again $dz=-vdt$)
\EQN{7}{1}{}{}{ \RDLAST{ \CELL{\frac{dW}{dz}&= &-2\pi \epsilon
_0cd\int_{-\infty }^{+\infty }dtE_z\left( t\right) B_y\left(
t\right) }}{1}{}{}{} {}}
Converting this time integral into a frequency integral, we
obtain
\EQN{7}{1}{}{}{ \RDLAST{ \CELL{\frac{dW}{dz}&= &-2\epsilon
_0cd\,\text{Real\thinspace }\int_0^{+\infty }d\omega E_z\left(
\omega \right) B_y^{\ast }\left( \omega \right) }}{1}{}{}{} {}}
where $B_y\left( \omega \right) =-\partial A_z\left( \omega
\right) /\partial d$. For a point charge $q$,  $\Phi \left(
\omega \right) $ is given by (\ref{phiomc}) and $E_z\left( \omega
\right) $ by (\ref{Ezc}).  Using the expansions for large
arguments of the Hankel functions $H_0^1$ and $H_1^1$, we obtain
the power radiated by the particle
\EQN{7}{1}{}{}{ \RDLAST{ \CELL{\langle P\rangle &=
&-v\,\frac{dW}{dz} =  \frac{q^2v}{4\pi \epsilon
_0c^3}\int_0^{+\infty }d\omega \omega \left(
1-\frac{c^2}{n^2v^2}\right) }}{1}{}{}{} {}}
which is the well-known Frank-Tamm result. If instead of a point
charge, the charge is  distributed round a circle of radius $a$,
we have from Eqs. (\ref{Aomc}), (\ref{phiac}), (\ref{F2c}), and
(\ref{E2cz})
\EQN{7}{1}{}{}{ \RDLAST{ \CELL{\langle P\rangle &=
&\frac{q^2v}{4\pi \epsilon _0c^3}\int_0^{+\infty }d\omega \omega
\left( 1-\frac{c^2}{n^2v^2}\right) \left[ J_0(\omega a/\gamma
_cv)\right] ^2\label{Pc}}}{1}{}{}{} {}}
with $\gamma _c$ given by (\ref{gammacc}) - a result that can also
 be found directly using for example the general expression
derived by  Meyer-Vernet   (\cite{mey88}) with a different
formulation.

The total electric field is thus much greater than would be
suggested from an order of magnitude estimate using (\ref{Pc}).
Indeed (\ref{Pc}) involves the large distance values of $E_z$ and
$B_y=-vn^2E_x$, whereas the antenna measures essentially a local
value of $E_x$, which is much greater than $E_z$. 

\section{Discussion}

\subsection{Physical significance of the results}

We have calculated analytically the boosted Coulomb and 
\v{C}erenkov  contributions to the electric field spectrum
produced by an extensive cosmic ray shower, with a highly
simplified model. These field contributions are due to the
relativistic speed of the radiating charges and do not depend on
the acceleration. The result may be understood more intuitively
by using  Feynman's formula $E=-q/\left( 4\pi \epsilon
_0n^2c^2\right) d^2\theta /dt^2 \times$sign($1-n\BFS{\beta}\cdot
\BF{n}$),  where  $\theta $ 
is the retarded angular position of the particle (Fig.
\ref{geom1}). This formula, which can be derived from the 
Li\'{e}nard-Wiechert potentials used in the present paper
(Feynman \cite{fey64}), was used by Allan (\cite{all71}) in his
seminal review.  With Allan's notations (origin at $t=0$ when the
charge passes at closest approach C), we have from Fig.
\ref{geom1}: $t=d/\left( c\sin \theta /n\right) -d/\left( v\tan
\theta \right) $.  The solutions for $\theta $  (one solution for
$v<c/n$, two solutions in the \v{C}erenkov  cone for  $v>c/n$),
and the second derivatives are given respectively in
Eqs.(A16)-(A17) and  Fig. A5 of the paper by Allan
(\cite{all71}). This yields an electric field varying as $\left(
1+t^2/\tau ^2\right) ^{-3/2}$ for $v<c/n$  and as $-H\left(
t-\tau \right) \left( t^2/\tau ^2-1\right) ^{-3/2}$ for $v>c/n$
($H$ being the Heaviside function), in agreement with Eqs.
(\ref{Ext}) and (\ref{Extc}) of the present paper. Feynman's
formula is more intuitive than the  Li\'{e}nard-Wiechert
formulation used here, but  it leads to more complicated
calculations for the present problem.

Feynman's formula enlightens the fact that the boosted Coulomb
and  \v{C}erenkov fields both stem from the strong acceleration
of the retarded angular trajectory at relativistic speeds, even
when the charge's speed is a constant. This takes place during a
short interval  $\tau $ of time near the instant when the
charge's path (not necessarily the charge itself as we already
noted, see Feynman \cite{fey64}) passes at closest approach to
the antenna (if $v<c/n$) or has the antenna on its \v{C}erenkov
cone (if $v>c/n$). As already noted, the boosted Coulomb and the 
\v{C}erenkov field have the same spectrum at low frequencies
($2\pi f<1/\tau $), which is the time integral of the  field, and
is independent on both the Lorentz factor  $\gamma $ and the
refractive index $n$.

\subsection{Boosted Coulomb versus \v{C}erenkov}

An important fact about cosmic ray showers is that the median
speed of secondary charged particles is roughly equal to the
phase speed of radio waves in air at about half the atmosphere
scale height. As a consequence, roughly the lower half of the
particle energy distribution produces a boosted Coulomb field,
roughly the higher half  produces a \v{C}erenkov field, whereas
the particles moving nearly at the wave phase speed should not
contribute except at very small distances, because in that case
the retarded altitude of emission (respectively (\ref{hret}) and
(\ref{hretc}) for the Coulomb and Cerenkov cases) jumps above the
atmosphere.

The boosted Coulomb and \v{C}erenkov field spectra of a point
charge are equal at small distances and/or frequencies, but the
Coulomb field decreases with frequency and distance more rapidly
than the \v{C}erenkov one. However,  the \v{C}erenkov field at
large frequencies or distances is produced when the antenna is
very close to the  \v{C}erenkov cone, so that it should be very
sensitive to the variation of the refractive index with altitude,
much more so than the boosted Coulomb field which does not
involve any singularity. Furthermore, the   \v{C}erenkov field 
is produced by the high energy tail of the shower particles, for
which the net charge and the magnetic separation are smaller. The
\v{C}erenkov field is thus expected to be smaller than the
boosted Coulomb field - a question that should be examined  more
carefully.

Finally, we have not considered the transition radiation (see
Ginzburg \&  Tsytovich   \cite{gin79}) that may be produced by
the variation in refractive index, especially when the charges
hit the ground.

\subsection{Comparison with synchrotron emission and with
observations}

The present estimate does not take into account the electric
field produced by the acceleration due to the Lorentz force (see
Huege \& al. \cite{hue07} and refs. therein), which yields 
opposite (time varying) deviations of the velocities of charges
of opposite signs. The field due to the acceleration by the
Lorentz force has been calculated as a special case of
synchrotron radiation (see Huege \& Falcke  \cite{hue03}) with
each particle completing only a small fraction of gyration since
the free path is much smaller than the gyroradius.

However, it should be noted that the published simulations of
that emission neglect the effect of refraction. Taking into
account the air refractive index may change the spectral density
and the radiation pattern, even for particles satisfying $n\beta
<1$, by introducing an ``equivalent Lorentz factor'' according to
(\ref{gammap}), whereas for $n\beta >1$, the charges can catch up
with the waves they emit. A naive application of the classical
expression of the synchrotron formulae (Jackson \cite{jac99})
would suggest that for a given charge, refraction should not
change significantly the field at small frequencies or distances
since in that case it does not depend on $\gamma $. However,
refraction should change the retarded altitude of emission, and
therefore the number of radiating charges, since it depends on
the development of the shower. In particular, charges with
$n\beta \simeq 1$ have a retarded altitude above the atmosphere
(except for very short perpendicular distances), and thus should
not contribute to radiation.

Nevertheless, it may be interesting to compare the boosted
Coulomb field with the published simulations of synchrotron
radiation. We find a Coulomb field strength (see for example Fig.
\ref{NKGdEWcf}) that may not be negligible compared to the values
calculated  for synchrotron radiation (see Huege \& al.
\cite{hue07} and refs. therein). Furthermore it appears to have a
 spectral  decrease with frequency and distance that is grossly
similar; for example the scale of decrease is of the order of
magnitude of 100 m at 20 MHz. More precise comparisons require
further studies which are outside the scope of this paper. Note
however that the boosted Coulomb and the synchrotron spectral
shapes  have different physical origins. The roughly exponential
decrease with both frequency and lateral distance is a generic
property of the boosted Coulomb field, which stems from the
expansion of the Bessel function $K_1$  for large arguments
(\ref{Ehf}), whereas the field strength tends to a finite
constant (\ref{Elf}) as $\omega \rightarrow 0$.  In contrast, the
synchrotron electric field spectrum of a point charge does
increase with frequency, with the field vanishing for  $\omega
\rightarrow 0$, so that the field decrease with frequency
exhibited by the synchrotron simulations is produced by the
longitudinal extension of the source and is very sensitive to the
time of arrival distribution (see Huege \& Falcke \cite{hue03}).

The boosted Coulomb field strength shown in Fig.  \ref{NKGdEWcf}
is of the same order of magnitude as the published observations
(see Allan \cite{all71}, Ardouin  \cite{ard06}, Horneffer
\cite{hor06}, Lecacheux \& al. \cite{lec07}), including the scale
of variation with distance. For example Fig. \ref{NKGdEWcf},
calculated with a plausible source size and a total number of
particles (roughly proportional to the primary energy)
corresponding to shower maximum development for a primary of 
$2\times 10^{17}$eV, yields a low-frequency field Fourier
transform of about 6 $\mu $V/m/MHz at 100 m perpendicular
distance, decreasing with a spatial scale of about 100 m; with
the same parameters, the relative excess $\eta $ of electrons
over positrons yields an additional field of $\eta $ times the
value in Fig. \ref{NKG}. Note that the spectral densities
(defined for positive frequencies only) are twice these values,
and that the above estimates are only order-of-magnitude ones.
Comparing with Fig. 38 of   Huege \& Falcke \cite{hue03} for
example suggests that this contribution may not be negligible.

Finally, since the boosted Coulomb field is produced by  the
charges satisfying $n\beta <1$ and whose retarded altitude
(\ref{hret}) (depending on $n$, too) is in the region of
significant shower development, it  depends on the refractive
index. The field should  thus be sensitive to atmospheric
pressure, temperature, and humidity.

\subsection{Defects of the model}

In order to obtain analytical results, we made many
simplifications which make our work little more than a
preliminary step for more detailed calculations. In particular,
one should take into account the variation of the refractive
index with altitude, the longitudinal distribution of the source,
the variation of the charge distribution with (retarded) 
altitude, the deviations of the particle velocities, and the 
particle velocity distribution. Such studies, which require a
detailed numerical simulation of the shower, are outside the
scope of this paper.

In particular, as indicated in Section 4, the longitudinal extent
of the source should reduce the amplitude above about 50 kHz, 
because of loss of coherence.

In order to obtain analytical results, we have considered a fixed
value of $\gamma $, close to the median value for shower
secondaries. A realistic calculation should involve an
integration over the particle velocity distribution, taking care
of   the singularity at  $n\beta =1$. This singularity is
expected to be alleviated by the fact that the corresponding
retarded altitude jumps above the atmosphere, so that the charges
producing it cannot contribute. Hence, at distances of  hundreds
of metres and frequencies of tens of MHz, the integration of the
field strength over $\gamma $ is expected to produce a result
behaving roughly as calculated in Sections 3 and 4, with a  value
of $\gamma ^\prime $ roughly equal to the maximum value for which
the retarded altitude yields a significant charge.

\section{Conclusions}

We have evaluated the electric field produced by the relativistic
velocity of the charges in extensive cosmic ray showers, using a
highly simplified model of the charge distribution in order to
obtain analytical results. We show that the refractive index
plays an important role, not only in producing a
\v{C}erenkov-like field at supraluminal speeds, but also in
changing the emission at subluminal speeds since the bulk of
secondary particles move roughly at the wave phase speed. In
particular, except for small perpendicular distances, this puts
above the atmosphere the retarded altitude of the particles
moving close to the wave phase speed, so that they do not
contribute to the field. Even though our calculations are based
on oversimplified hypotheses, the analytical formulas obtained
are useful to estimate the importance of various parameters and
physical processes for the part of the field that does not depend
on the acceleration. Our main results are listed below.

\noindent 1) A little less than the lower half of the charge
energy distribution in the shower produce a boosted Coulomb
electric field. With plausible parameters of the charge distribution,
this yields an electric field spectrum comparable to the values
calculated for synchrotron radiation and to those observed. In
particular, we find that a primary of $10^{17}$ eV should produce
a spectral field of a few $\mu $V/m/MHz  at 20 MHz and 100 m
perpendicular distance, decreasing roughly as an exponential with
the product of  perpendicular distance by  frequency, with a
scale of decrease of about 100 m at 20 MHz. This is comparable to
current measurements, and therefore does not appear to be
negligible.

\noindent 2) Charges of higher energy produce a \v{C}erenkov-like
field. However, this field may be more difficult to detect for
several reasons: first, the relative excess of electrons over
positrons is small at these energies, making the net charge
small; second, the magnetic separation between electrons and
positrons decreases with energy as $1/\gamma $; third, the phase
varies with lateral position, so that the finite \itshape
lateral\upshape\hspace{0pt} size of the source decreases the
field amplitude because of loss of coherence, except at very
small frequencies; this is not the case of the Coulomb field, for
which loss of coherence is only produced by the (small) \itshape
longitudinal\upshape\hspace{0pt} extension. And finally, it is
only at large frequencies or distances ($\omega d/\gamma
_c^\prime v\gg 1$) that the  \v{C}erenkov-like field may be
greater than the boosted Coulomb contribution, and in this range
the variation of the refractive index with altitude should play
an important role.

\noindent 3) The radio electric field depends on the air
refractive index, and thus on the pressure, temperature and
concentration of various components, especially humidity. This
should introduce variations of radio emission with the altitude
of emission and reception, and may produce significant day-night,
seasonal and other effects.

\noindent 4) This study suggests several developments which are
outside the scope of this paper. In particular, taking into
account the variation of the refractive index with altitude, and 
 making a detailed comparison with synchrotron emission.

\begin{acknowledgements}
      D. Ardouin is grateful to Observatoire de Paris for hospitality and support.
\end{acknowledgements}

\end{document}